\documentclass[%
 aip,
 amsmath,amssymb,
 reprint,%
]{revtex4-1}

\usepackage{graphicx}
\usepackage{dcolumn}
\usepackage{bm}
\usepackage[utf8]{inputenc}
\usepackage[T1]{fontenc}
\usepackage{mathptmx}
\usepackage{etoolbox}
\usepackage{siunitx}
\usepackage{gensymb}

\makeatletter
\def\@email#1#2{%
 \endgroup
 \patchcmd{\titleblock@produce}
  {\frontmatter@RRAPformat}
  {\frontmatter@RRAPformat{\produce@RRAP{*#1\href{mailto:#2}{#2}}}\frontmatter@RRAPformat}
  {}{}
}%
\makeatother
\begin{document}

\preprint{AIP/123-QED}

\title[Disorder-driven transport in IBAS ZrN films]{Disorder-driven transport in ZrN thin films\\ grown by ion-beam-assisted sputtering}

\author{M. DiPreta}
\affiliation{Physics Division, Argonne National Laboratory}
\affiliation{Materials Science Division, Argonne National Laboratory}

\author{S. Jiang}
\affiliation{Materials Science Division, Argonne National Laboratory}

\author{S. Kruhlov}
\affiliation{Drexel University, Physics Department}
\affiliation{Physics Division, Argonne National Laboratory}

\author{Y. Li}
\affiliation{Materials Science Division, Argonne National Laboratory}

\author{V. Novosad}
\affiliation{Materials Science Division, Argonne National Laboratory}
\affiliation{Physics Division, Argonne National Laboratory}

\author{T. Polakovic}
\email[Corresponding author: ]{tpolakovic@anl.gov}
\affiliation{Physics Division, Argonne National Laboratory}

\date{\today}

\begin{abstract}

We investigate the structural and electronic transport properties of zirconium nitride (ZrN) thin films grown by dc ion-beam assisted sputtering (IBAS) as a function of nitrogen partial pressure, sputtering power and deposition temperature. Variation of growth conditions enables controlled tuning from disordered metallic and superconducting behavior to insulating transport. Increasing nitrogen flow drives an increase in sheet resistance, suppression of superconducting transition temperature, and non-metallic conduction. XRD shows no measurable change in long range crystallinity or order. In the insulating regime, low-temperature transport is well described by 3D Mott variable-range hopping (VRH), with the characteristic temperature $T_0$ increasing monotonically with sheet resistance over four orders of magnitude. Despite this strong correlation between $T_0$ and disorder, $T_c$ exhibits no systematic dependence on $T_0$, indicating a decoupling between localization physics and the superconducting energy scale. Magnetotransport measurements are not well described by the standard BCS model nor the dirty type-II Werthamer--Helfand--Hohenberg model. The phase boundary is instead captured by a free-exponent power law, $\mu_0 H_{c2}(T) = \mu_0 H_{c2}(0)\left[1 - (T/T_c)^{n}\right]$ with $n \approx 3.47$,and extrapolates to $\mu_0 H_{c2}(0) \approx 6.4$~T. The films have an extracted coherence length of \qty{7}{nm}. These results establish IBAS-grown ZrN as a broadly tunable platform for investigating disorder-driven transport and the crossover between metallic conduction, electronic localization, and superconductivity.
\end{abstract}

\maketitle

\section{Introduction}
Group IV transition metal nitrides (TMNs) have been used throughout various industries such as nuclear reactors and drill bit coatings, due to their characteristically high melting points, hardness, and chemical stability.\cite{ZrN_mech1, ZrN_mech2, ZrN_mech3} Applications have broadened beyond mechanical uses and are now being explored. Titanium nitride has received particular attention due to its metallic optical response, high carrier density, compatibility with CMOS processing, and superconducting properties,\cite{TiN, TiN2, TiN3} while zirconium nitride and hafnium nitride have received comparatively less attention despite sharing similar structural and electronic characteristics.

Zirconium nitride has attracted increasing interest as a functional electronic material, finding use both as a conductive material in photonic, plasmonic, and CMOS-compatible devices\cite{ZrN_con_1, ZrN_con_2, ZrN_con_3, ZrN_con_4} and as a dielectric or insulating layer in electronic and microelectronic applications.\cite{ZrN_die_1, ZrN_die_2, ZrN_die_3, ZrN_die_4} Evidently, ZrN provides a tunable platform for investigating electronic transport across conductive, resistive, and dielectric regimes. This broadband tunability makes it an attractive material to study quantum phase transitions, such as the superconductor–insulator transition, where quantum fluctuations dominate behavior. 

Ion-beam assisted sputtering (IBAS) has been shown to enhance film densification and adhesion, reduce void formation, and improve crystalline texture compared to conventional reactive magnetron sputtering.\cite{IB1,IB2,IB3} More recently, IBAS has been reported to increase the superconducting transition temperature ($T_c$) of NbN and TiN films deposited at room temperature.\cite{NbN,TiN}

In IBAS, the reactive gas is introduced through an ion source, where molecular nitrogen ($N_2$) is ionized and accelerated toward the substrate. Near the exit of the ion gun, the accelerated ions are neutralized through interaction with an electron beam, producing high-energy neutral atomic nitrogen. These energetic species bombard the growing film surface, enhancing surface mobility and reaction kinetics, thereby promoting improved microstructure and, therefore, superconducting properties.

In this work, we investigate the electronic transport properties of IBAS grown ZrN thin films and establish their relationship to synthesis conditions, crystallographic and micro structure, magnetic field response, and temperature. Through variation of nitrogen partial pressure and deposition temperature, we explore metallic, superconducting and variable range hopping (VRH) regimes. Through the study of crystallographic evolution alongside transport behavior under varying temperature and magnetic field, we gain insight into the role of disorder in governing the crossover between metallic conduction, electronic localization, and superconductivity in IBAS grown ZrN.

\section{Methods}

ZrN films were grown on high-resistivity ($>\qty{10}{\kilo\ohm}$) (100) Si wafers. All films were deposited using an IBAS process in a commercial ultra-high vacuum sputtering system (Angstrom Engineering). Wafers were introduced through a load-lock, and the chamber was pumped down to a base pressure of $1 \times 10^{-7}$ Torr prior to deposition.

A \SI{3}{in} Zr target (\qty{99.95}{\percent} purity) was used for all depositions. The argon flow (\qty{99.9999}{\percent} purity) was fixed at \qty{30}{sccm}, while the nitrogen flow ($N_2$, \qty{99.9997}{\percent} purity) was varied. The deposition pressure was maintained at \SI{3}{mTorr}, and the substrate temperature for some samples was varied. Prior to growth, the substrate surface was subjected to a low energy Ar ion-beam to etch away residual moisture or organic contamination. Additionally, prior to deposition the Zr target was pre-sputtered in an Ar atmosphere for \SI{10}{\minute} to remove surface contamination. Nitrogen was then introduced via the ion gun, and the Zr target was sputtered for an additional \SI{10}{\minute} to promote target surface nitridation. 

Depositions were carried out at power densities ranging from \SI{3.18}{\watt\per\centi\meter\squared} to \SI{4.82}{\watt\per\centi\meter\squared}. All films had a thickness of \SI{200}{nm}, verified ex-situ with a profilometer. 

The substrate-to-target distance was fixed at \SI{5}{in}, with the target oriented 33$^\circ$ relative to the substrate normal. An end-Hall ion source equipped with a hollow cathode was used to generate and neutralize the nitrogen ion beam. The ion source was positioned at 40$^\circ$ with respect to the substrate and offset azimuthally by 20$^\circ$ from the Zr target.

During IBAS, nitrogen was supplied exclusively through the ion source rather than uniformly throughout the chamber. The ion energy was limited to \SI{100}{eV} to minimize ion-induced damage and suppress microstructural defects. An ion current of \SI{0.5}{A} was maintained during deposition, corresponding to an ion power density of approximately \SI{70}{mW\,cm^{-2}}. The general deposition parameters are summarized in Table~\ref{tab:dep_param}.

\begin{table}[htbp]
\centering
\caption{Summary of ZrN deposition parameters}
\label{tab:dep_param}
\begin{tabular}{ll}
\hline
\textbf{Parameter} & \textbf{Value / Description} \\
\hline
Substrate & Si (100)($>\qty{10}{\kilo\ohm}$)  \\
Base pressure & $1 \times 10^{-7}$ Torr \\
Target & \SI{3}{in} Zr (\qty{99.95}{\percent} purity) \\
Ar flow & \qty{30}{sccm} \\
$N_2$ flow & Variable \\
Deposition pressure & \SI{3}{mTorr} \\
Deposition temperature & \SIrange[]{300}{500}{C} \\
Power density & \SIrange{2.41}{4.82}{\watt\per\centi\meter\squared} \\
Film thickness & \SI{200}{nm} \\
Substrate–target distance & \SI{5}{in} \\
Ion energy & \SI{100}{eV} \\
Ion current & \SI{0.5}{A} \\
\hline
\end{tabular}
\end{table}

Transport properties were measured using the standard four-probe technique in a Quantum Design, Inc. PPMS. Prior to electrical characterization, aluminum contact pads were deposited onto the ZrN surface to minimize contact variability and ensure stable electrical contacts during transport measurements. Most films were measured from \SIrange[]{300}{3}{K}, with some being measured down to \SI{100}{mK}. Samples were diced along the (100) Si plane and imaged with an SEM in cross sectional view. 

\section{Results \& Discussion}

\begin{figure}[!t]
    \centering
    \includegraphics[width=1\linewidth]{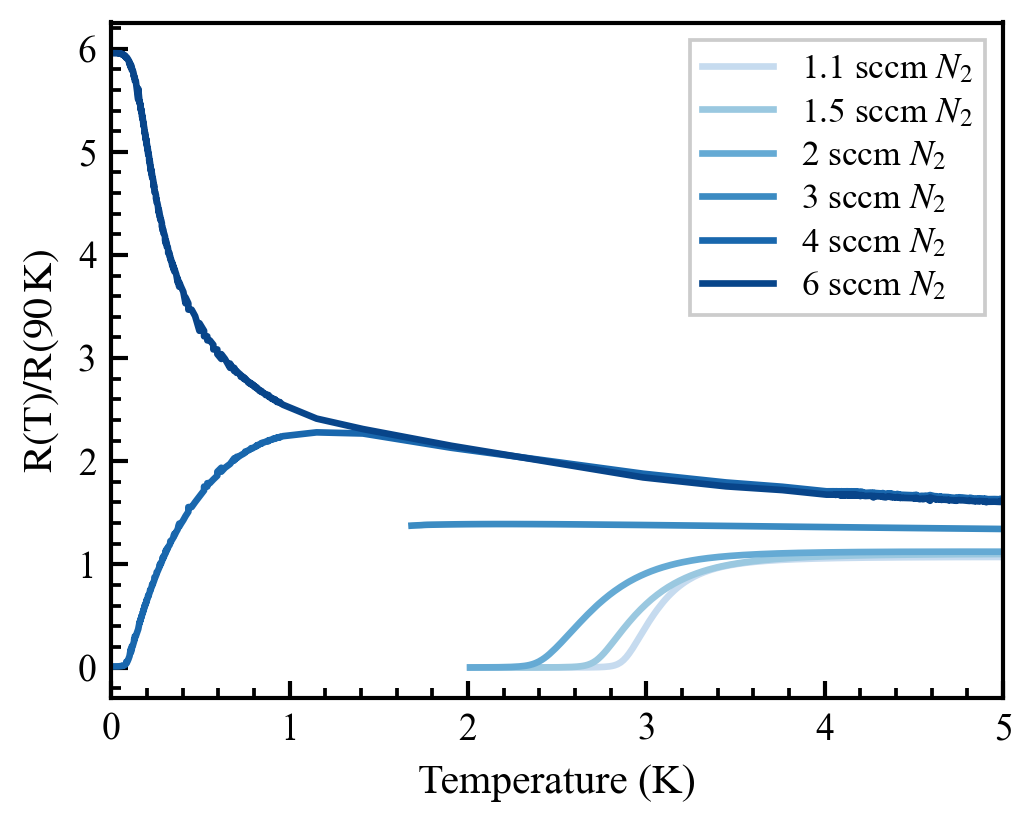}
    \caption{\label{fig:RT} Normalized resistance, $R_T/R_{\SI{90}{\kelvin}}$, as a function of $\mathrm{N_2}$ flow rates from \SIrange{1.1}{6.0}{sccm} for \SI{200}{nm} ZrN thin films. The superconducting transition temperature, $T_c$, is progressively suppressed and the transition broadens with increasing nitrogen flow. The \SI{3.0}{sccm} sample shows the onset of a transition but was not measured below the cryostat base temperature of \SI{1.68}{\kelvin}.}
\end{figure}

As the first important broad result, we identify the trend of increased nitrogen flow leading to more resistive, insulating behavior accompanied by a suppression of T$_c$, consistent with established reports of disorder-driven transport evolution in nitride films \cite{amorphous_1, amorphous_2, amorphous_3, amorphous_4}. This behavior is captured in Fig. \ref{fig:RT}, where increasing nitrogen flow leads towards more resistive behavior and lower T$_c$ of the films. Because electronic disorder, reflected in sheet resistance, is more consistent at quantifying the resulting film state, samples are organized and compared using electrical transport metrics.

\begin{figure}[!b]
    \centering
    \includegraphics[width=0.8\linewidth]{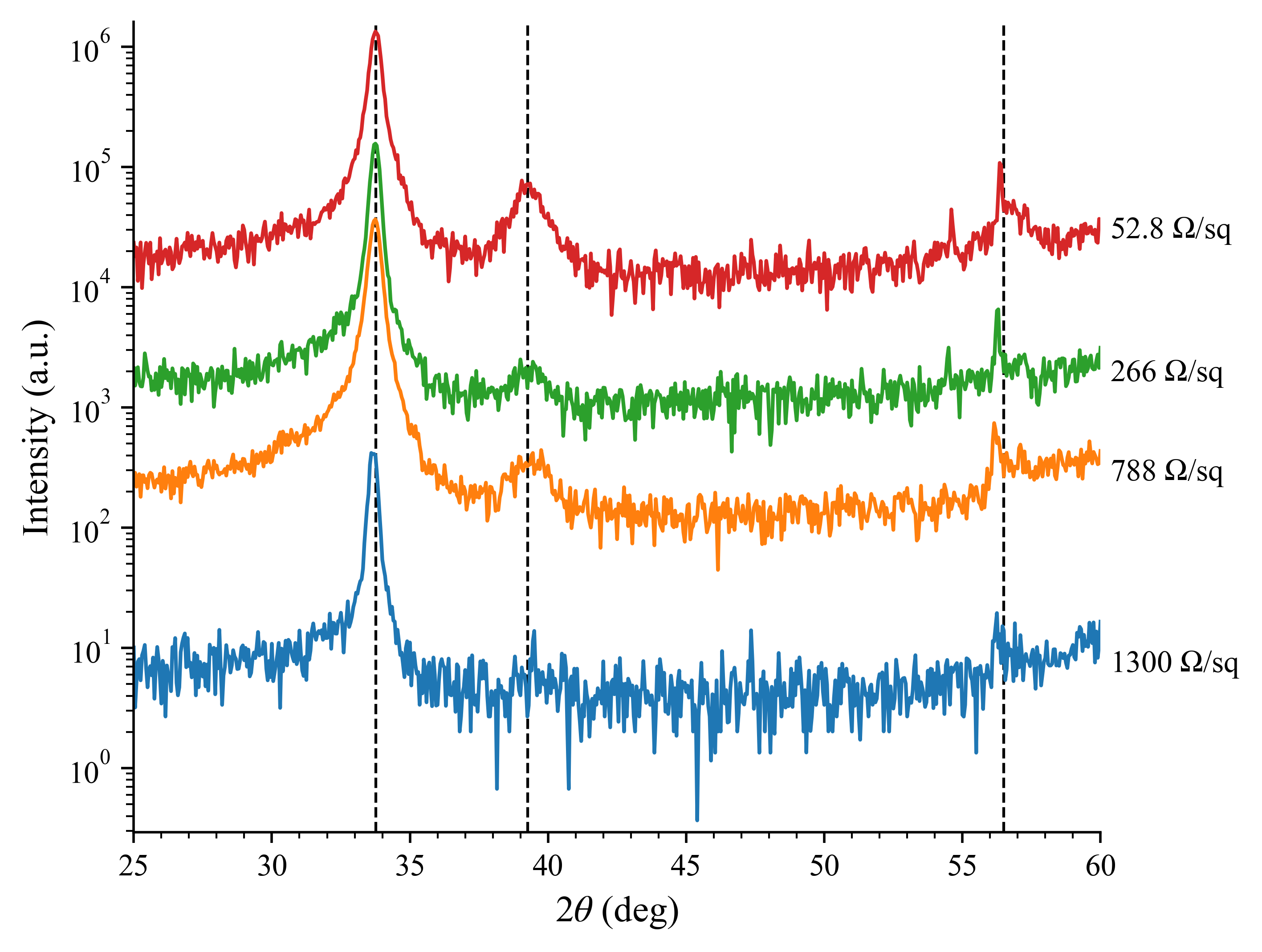}
    \caption{\label{fig:xrd} XRD of ZrN thin films with varying $R_\square$. Dashed lines indicate the expected diffraction positions of the cubic ZrN (111), (200), and (220) reflections, located at $33.8^\circ$, $39.2^\circ$, and $56.6^\circ$, $2\theta$ respectively. The $788\ohm/\square$ sample is superconducting at \SI{3.2}{K}.} 
\end{figure}

Ion-beam assisted sputtering (IBAS) has previously been demonstrated to enhance film texture in NbN and TiN,\cite{NbN, TiN} and thus was chosen as the deposition method for ZrN. The IBAS grown ZrN films, spanning almost three orders of magnitude in sheet resistance, exhibit little change in crystallinity. As $R_\square$ increases, the (111) reflection shifts toward lower diffraction angles, while the secondary peaks such as (200) can be seen increasing in intensity as $R_\square$ decreases. The shift to lower angle corresponds to an increase in lattice spacing, consistent with increasing structural disorder. 

Prior reports show that elevated nitrogen partial pressure produces nitrogen-rich $\mathrm{ZrN}_x$ films with enhanced disorder and reduced crystalline coherence \cite{amorphous_1, amorphous_2, amorphous_3, amorphous_4}, however, XRD provides relatively little information to describe the change in $R_\square$ and transport measurements for IBAS grown ZrN samples. 
\begin{figure}[t]
    \centering
    
    \includegraphics[width=0.5\textwidth]{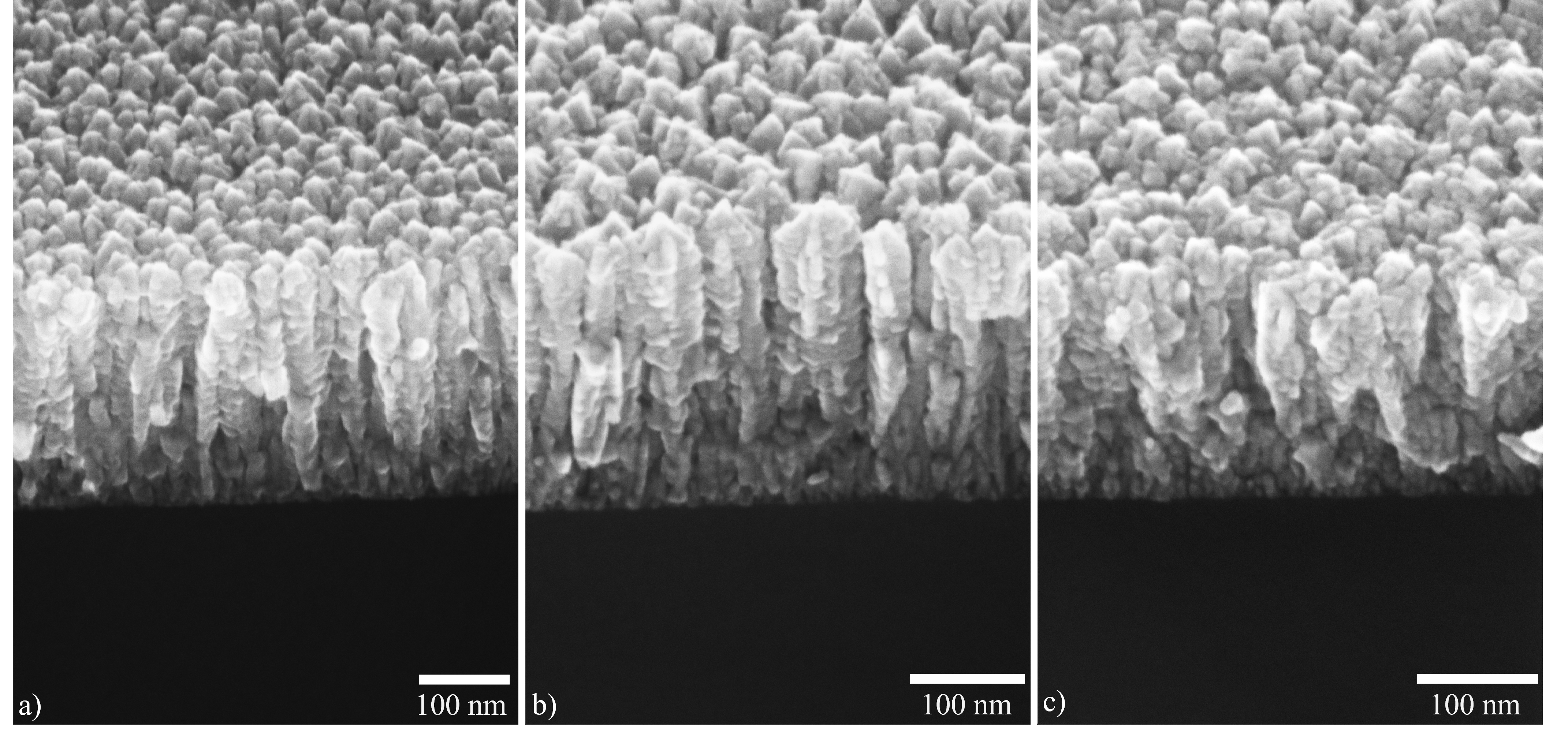}
    
    \vspace{0.25cm}
    \includegraphics[width=0.5\textwidth]{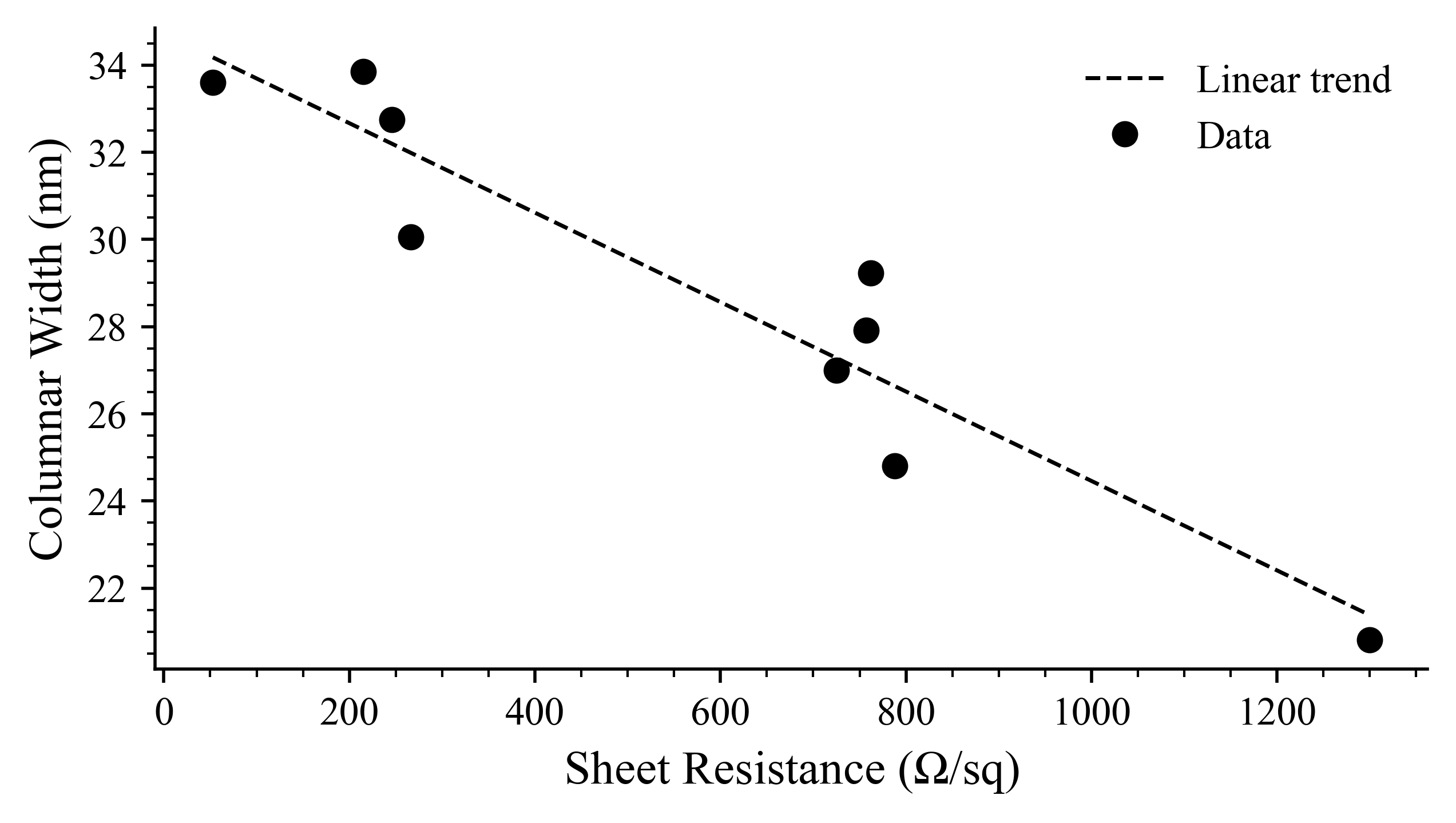}
    
    \caption{(a--c) Cross-section SEM images of 200-nm-thick IBAS ZrN films grown at room temperature via decreasing nitrogen flow, with sheet resistances of $747~\ohm/\square$, $215~\ohm/\square$, and $52.8~\ohm/\square$, respectively. (d) SEM-measured columnar grain width plotted against sheet resistance across various films, showing an inverse relationship consistent with grain-boundary-limited scattering.}
    \label{fig:combined}
\end{figure}

Upon further investigation of the film, a trend can be observed when comparing $R_\square$ to the columnar width of the ZrN films. The cross sectional SEM images shown in \ref{fig:combined}(a), (b) and (c) show an evolution of microstructure across various $R_\square$. Samples with smaller $R_\square$ have larger grains while samples with larger $R_\square$ values demonstrate more distinct columnar grains with smaller widths. A linear inverse relationship is identified as shown in Fig. \ref{fig:combined}(d).

A structure zone diagram is useful to illustrate the features of sputtered films and thus we will refer to the diagram depicted in Anders, 2009\cite{zones} to categorize the various microstructures observed. ZrN exhibits its highest T$_c$ with a well defined (111) phase in its cubic structure \cite{}, which usually manifests itself as columnar growth of this phase, as seen in other TMNs\cite{NbN, TiN}. $700\ohm/\square$--$800\ohm/\square$ in \ref{fig:combined}(a), show individual and distinct columnar structures with some recrystallization, which we best describe as being in zone T. As R$_\square$ values drop below $700\ohm/\square$ neighboring columns begin merging, showing signs of recrystallization \ref{fig:combined}(b), which could be characterized as a zone 2 structure. Samples with $R_\square$ measuring less than $200\ohm/\square$ in \ref{fig:combined}(c) demonstrate a structure akin to that in zone 3 with recrystallized grain structure with poor columnar formation. 
\begin{figure}[t]
    \centering
    \includegraphics[width=1\linewidth]{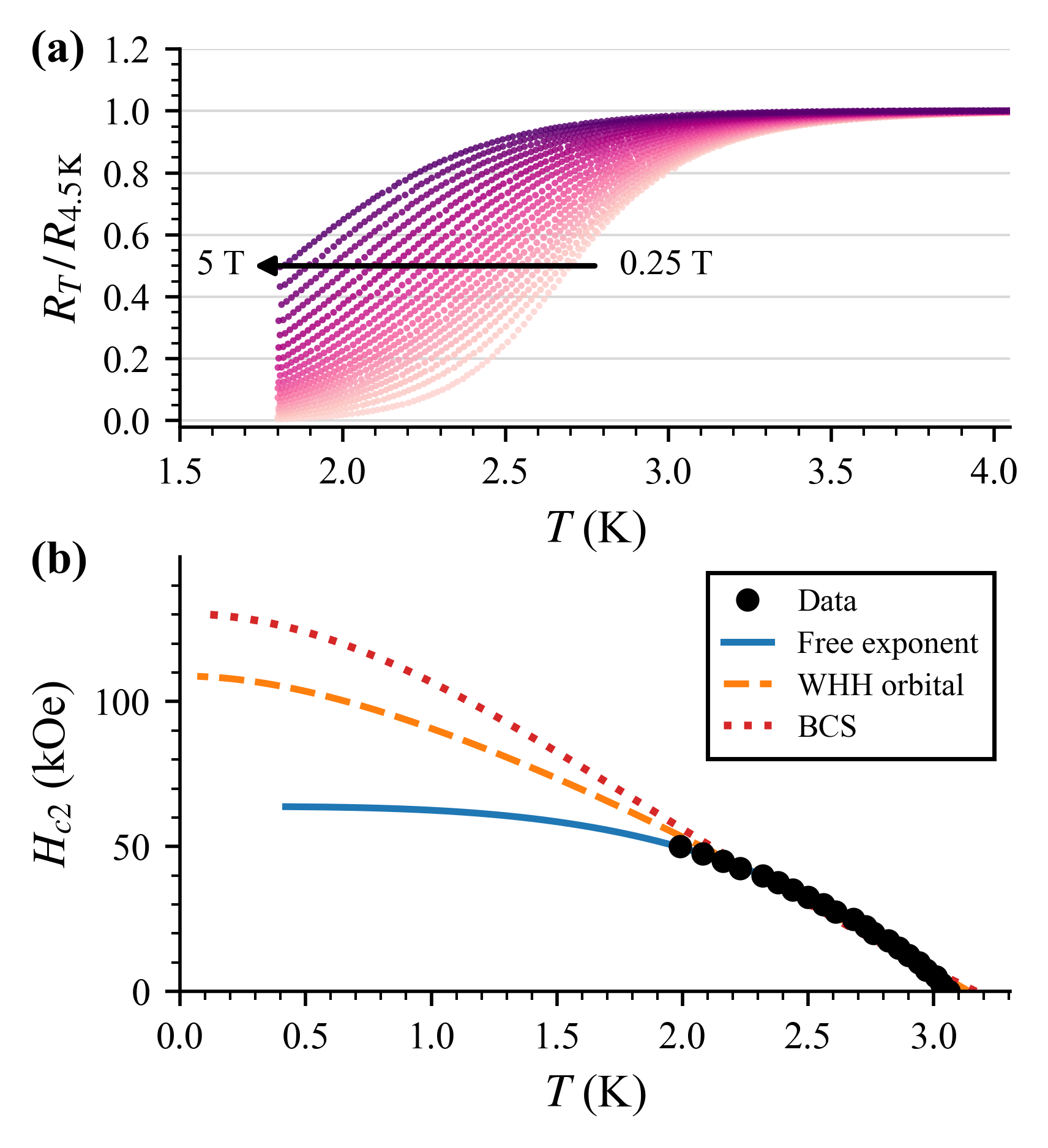}
    \caption{\label{fig:Hc} (a) Normalized resistance, $R_T/R_{\SI{4.5}{\kelvin}}$, as a function of temperature under applied magnetic fields from \SI{0.25}{\tesla} to \SI{5}{\tesla}. The superconducting transition is progressively suppressed and broadened with increasing field. (b) Upper critical field, $H_{c2}(T)$, extracted from the midpoint of the resistive transition at each field, together with fits to an empirical free-exponent power law (n = 3.47), the WHH dirty-limit model, and a BCS/Gorter--Casimir interpolation. The free-exponent fit best reproduces the measured curvature ($R^2=0.999$) and extrapolates to $\mu_0 H_{c2}(0) \approx \SI{6.4}{\tesla}$; the WHH and BCS forms extrapolate to higher values ($\approx\SI{10.9}{\tesla}$ and $\approx\SI{13.0}{\tesla}$, respectively), illustrating the model-dependence of the zero-temperature extrapolation.}
    
\end{figure}

Fig.~\ref{fig:Hc}a shows the systematic suppression of the superconducting transition with increasing magnetic field. The transition evolves continuously and exhibits the curvature expected for a dirty-limit type-II superconductor. We compare the extracted $H_{c2}(T)$ against three models. The first is the parabolic form
\begin{equation}
\label{eq:parabolic}
H(T) = H(0)\left[1 - \left(\frac{T}{T_c}\right)^2\right],
\end{equation}
which describes both the thermodynamic critical field of a type-I superconductor and the BCS/Gorter--Casimir interpolation.\cite{BCS1, GCI1} The second is the Werthamer--Helfand--Hohenberg (WHH) model for orbital pair breaking in the dirty limit \cite{WHH66, HW66},
\begin{equation}
\label{eq:whh}
\ln\!\left(\frac{1}{t}\right) = \psi\!\left(\frac{1}{2} + \frac{h}{2t}\right) - \psi\!\left(\frac{1}{2}\right),
\qquad
h = \frac{4H_{c2}}{\pi^2 T_c \left|dH_{c2}/dT\right|_{T_c}},
\end{equation}
where $t = T/T_c$ and $\psi$ is the digamma function. Unlike Eq.~\ref{eq:parabolic}, the WHH curve has no free shape parameter: it is fixed entirely by $T_c$ and the slope of $H_{c2}$ at $T_c$. The third is an empirical power law with a free exponent,
\begin{equation}
\label{eq:freeexp}
H_{c2}(T) = H_{c2}(0)\left[1 - \left(\frac{T}{T_c}\right)^{n}\right],
\end{equation}
which relaxes the fixed quadratic dependence of Eq.~\ref{eq:parabolic}.

The parabolic form (Eq.~\ref{eq:parabolic}) is the poorest description ($R^2 = 0.968$), deviating from the measured $H_{c2}$ at higher fields; this failure indicates that the films do not follow type-I scaling. The WHH model performs better ($R^2 = 0.983$), consistent with dirty-limit orbital pair breaking, while the free-exponent form gives the closest description of the data ($R^2 = 0.999$) with $n = 3.47$. We treat Eq.~\ref{eq:freeexp} only empirically as it carries an additional free parameter relative to Eq.~\ref{eq:whh}. 

\begin{figure}[b]
    \centering
    \includegraphics[width=1\linewidth]{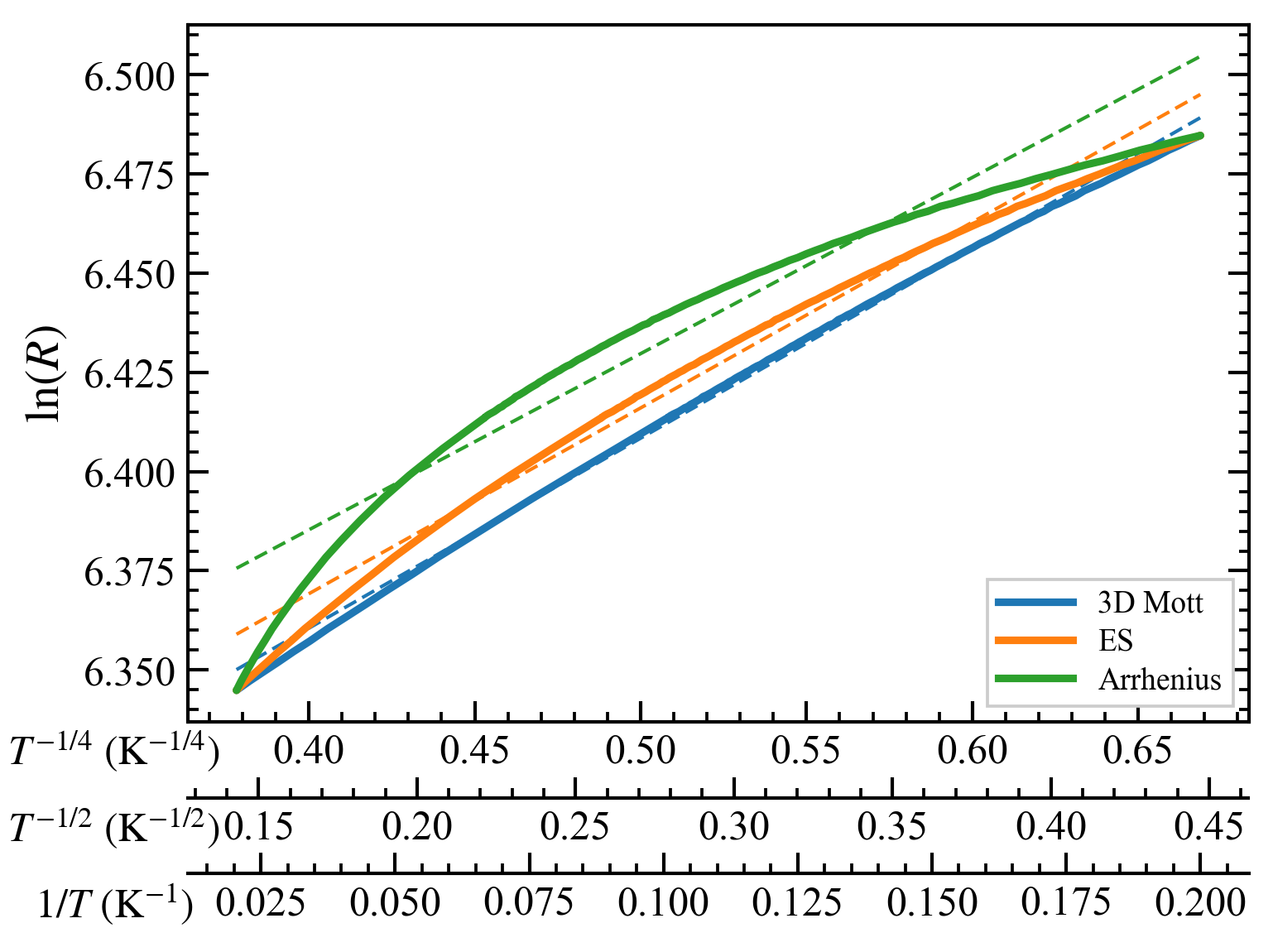}
    \caption{\label{fig:VRH} Linearized low-temperature transport (15–50 K) of a 200 nm ZrN film plotted as $\ln R$ versus $T^{-1/4}$ (3D Mott variable-range hopping), $T^{-1/2}$ (Efros–Shklovskii hopping), and $1/T$ (Arrhenius activation). The most extended linear regime is observed for $\ln R$ vs $T^{-1/4}$, indicating that 3D Mott VRH provides the best description of the transport in this temperature window.}
\end{figure}

Because all three fits are constrained only by data above \qty{1.8}{\kelvin}, none of them
is anchored near $T = 0$, and their extrapolated zero-temperature critical fields differ
substantially: the free-exponent fit yields $\mu_0 H_{c2}(0) \approx \SI{6.4}{\tesla}$,
while the WHH and BCS forms extrapolate to $\approx\SI{10.9}{\tesla}$ and
$\approx\SI{13.0}{\tesla}$, respectively. Measurements repeated in fields up to
\qty{9.8}{\tesla} gave similar results and were again best described by the free-exponent
form.
 
The superconducting coherence length $\xi(0)$ follows from the critical field as
\[
\xi(0) = \left[\frac{\Phi_0}{2\pi\mu_0 H_{c2}(0)}\right]^{1/2},
\]
where $\Phi_0$ is the magnetic flux quantum. We quote $\xi(0) \approx \SI{7}{\nano\meter}$ from the best-fitting form, in agreement with the \qty{7}{\nano\meter} reported for reactively sputtered ZrN.\cite{ZrN_con_2} Alternatively, the WHH and BCS forms instead giving \SI{5.5}{\nano\meter} and \SI{5.0}{\nano\meter}. This is several times smaller than the columnar widths of $\approx\SI{25}{\nano\meter}$ measured by SEM, indicating that superconductivity develops within individual columns rather than averaging across them.
 
\begin{figure}[b]
    \centering
    \includegraphics[width=1\linewidth]{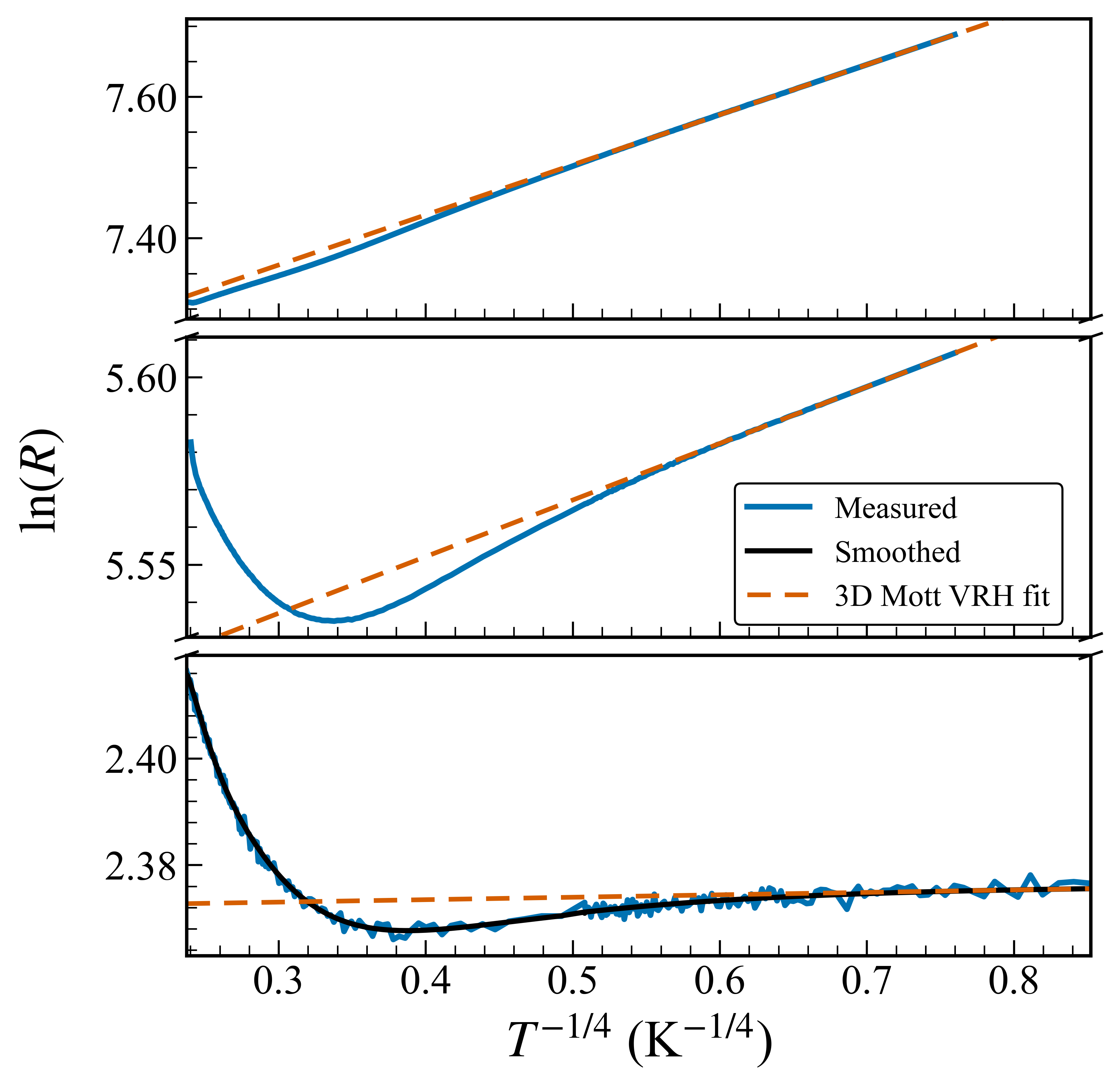} 
    \caption{\label{fig:var_temp} Linearized Mott variable-range hopping analysis of ZrN films deposited at different substrate temperatures. Deviations from linear behavior mark the onset of departure from Mott VRH transport. ZrN15 data was smoothed to improve fitting accuracy.}
\end{figure}

At higher temperatures, approximately 150--250~K, the resistance exhibits weak temperature dependence characteristic of a disordered metallic regime. In this regime, transport is dominated by extended electronic states, although strong elastic scattering suppresses conventional metallic behavior. The data in this temperature range are best described by a monotonic, metallic--like conduction model rather than activated or hopping--based transport.

Between approximately \qty{150}{K} \qty{50}{K}, the films undergo a smooth crossover from disordered metallic transport to hopping--dominated conduction. This crossover is marked by a gradual increase in resistivity and a breakdown of high--temperature transport models, rather than by a sharp transition. 

Below $\sim$\qty{50}{K}, the resistance is well described by three--dimensional Mott VRH, indicating phonon--assisted hopping between localized states. This window is the focus of the present analysis: at higher temperatures the films behave as disordered metals, while below $\sim$\qty{15}{K} the intrinsic hopping behavior is obscured by resistance saturation or the onset of superconductivity. Figure~\ref{fig:VRH} compares the linearized low--temperature data against 3D Mott, Efros--Shklovskii, and Arrhenius forms. The most extended linear regime is obtained for $\ln R$ versus $T^{-1/4}$, while the $T^{-1/2}$ and $1/T$ representations exhibit systematic curvature over the same interval. We therefore identify 3D Mott VRH as the dominant conduction mechanism between \qty{15}{K} and \qty{50}{K}.

At the lowest temperatures ($\lesssim 15$~K), several samples exhibit either a saturation of resistance or a transition into a superconducting state, signaling the onset of additional low--temperature transport mechanisms beyond the VRH regime.

The dominance of 3D Mott VRH is most pronounced in IBAS films deposited at room temperature. As the deposition temperature is increased, the temperature range over which VRH provides a good description of the transport  narrows. As shown in Fig.~\ref{fig:var_temp}, room--temperature deposited films exhibit good agreement with 3D Mott VRH down to $\sim 15$~K, while films deposited at \qty{250}{\celsius} deviate from VRH behavior below $\sim 8$~K. Films deposited at \qty{500}{\celsius} show little evidence of 3D Mott VRH above \qty{1.9}{K}, instead displaying predominantly metallic transport.

\begin{figure}[t]
    \centering
    \includegraphics[width=1\linewidth]{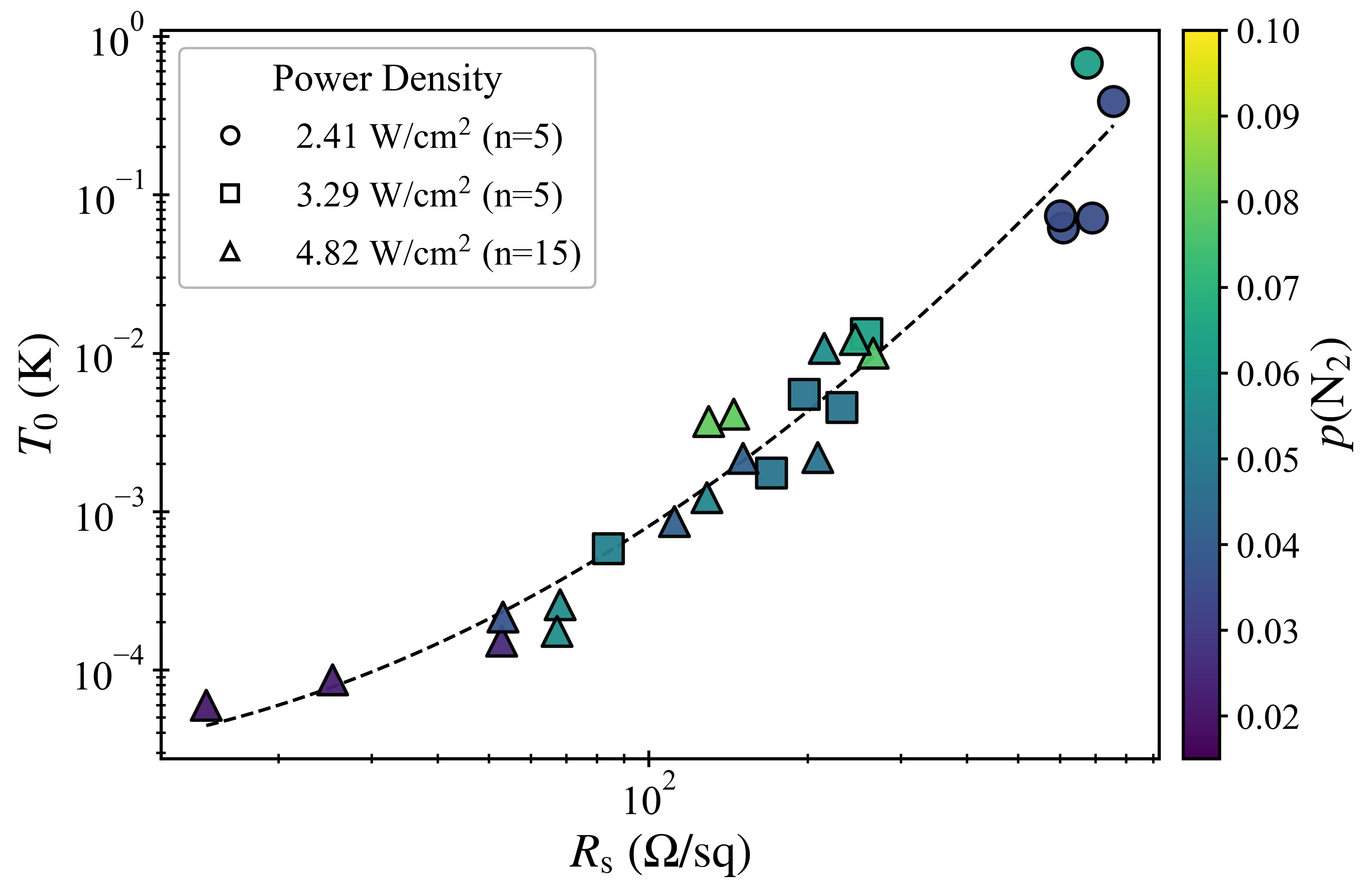}
    \caption{\label{fig:Rs} Characteristic Mott temperature $T_0$ extracted from 3D variable-range hopping fits (5--50 K) as a function of $R_\square$ for room-temperature sputtered ZrN films. Symbols denote sputtering power, and color indicates nitrogen partial pressure $p(\mathrm{N_2})$. The dashed line is a quadratic fit as a trend guide in log--log space.}
\end{figure}
This change indicates a reduction in disorder with increasing deposition temperature, consistent with prior growth studies in which lower deposition temperatures lead to more insulating behavior under otherwise comparable conditions.\cite{ZrN_e_prop2, Dome} A similar trend is observed with nitrogen content: reduced nitrogen incorporation favors metallic transport, while increased nitrogen flow drives the system toward non--metallic and insulating behavior as discussed earlier. In both cases, disorder emerges as the primary control parameter governing the crossover from metallic to localized transport in these films. 

Figure~\ref{fig:Rs} measures the the Mott VRH characteristic temperature $T_0$, extracted from low-temperature fits (15--50~K), against sheet resistance $R_{\square}$ or various films. $T_0$ increases monotonically with increasing $R_{\square}$. This shows a strong characteristic hopping energy scale and the degree of disorder in the films. Within the Mott VRH framework, larger values of $T_0$ correspond to reduced localization lengths, making the scaling with $R_{\square}$ consistent with disorder-driven l1
ocalization. Across the dataset, $T_0$ varies by nearly four orders of magnitude, from $\sim10^{-5}$ to $\sim10^{-1}$~K, demonstrating the broad tunability of the localization strength achieved in these systems. Superconducting samples fall on the same trend as non-superconducting films, indicating that superconductivity emerges from a common disorder-controlled electronic landscape rather than from a distinct transport regime.

\begin{figure}[t]
    \centering
    \includegraphics[width=1\linewidth]{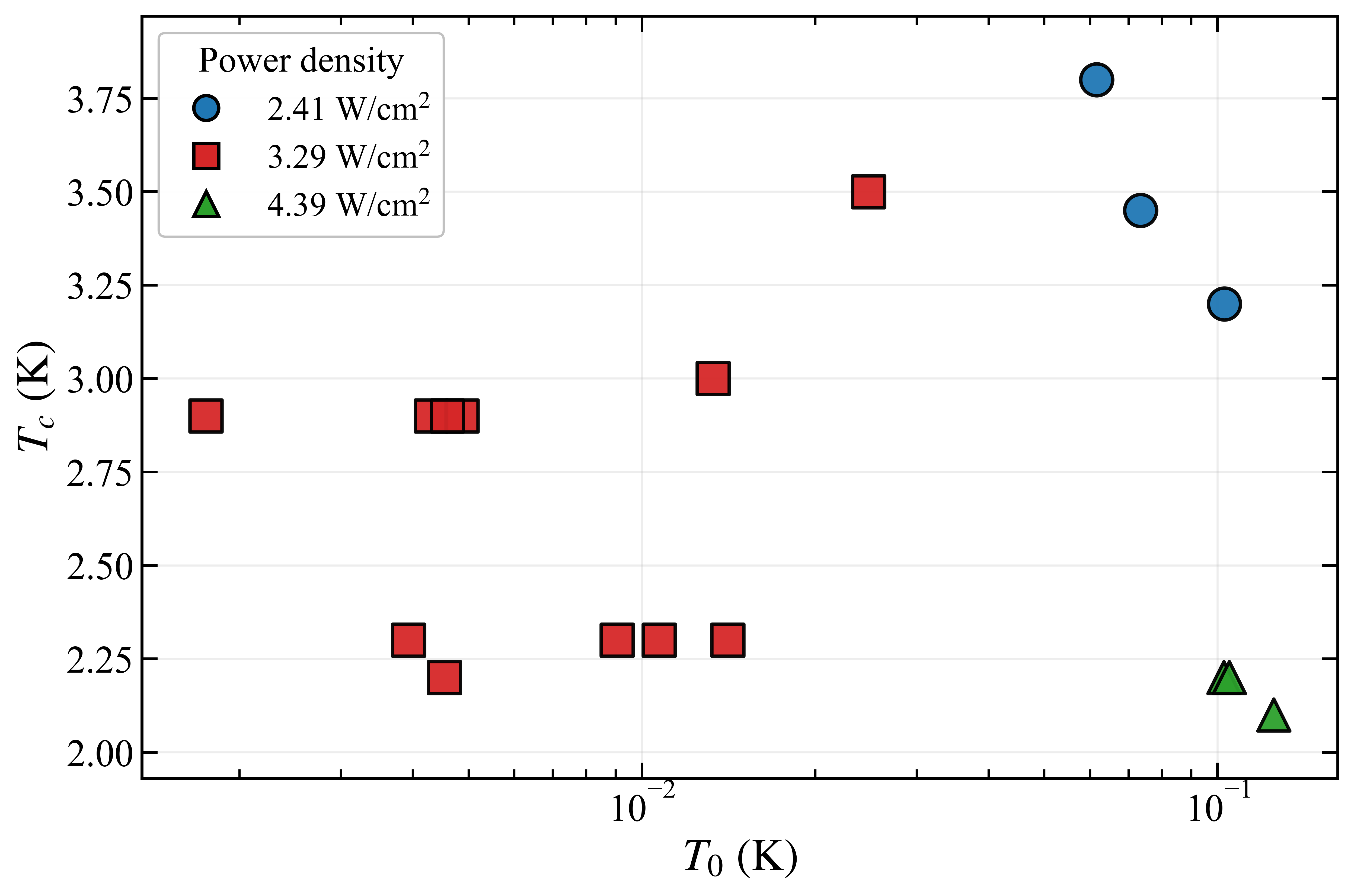}
    \caption{\label{fig:Tc} $T_c$ as a function of $T_0$ for ZrN films. Symbols denote sputtering power.}
\end{figure}

Figure~\ref{fig:Tc} shows the superconducting transition temperature $T_c$ plotted as a function of the Mott VRH characteristic temperature $T_0$, extracted from fits over the 15--50~K range. No clear dependence of $T_c$ on $T_0$ is observe as samples spanning nearly two orders of magnitude in $T_0$ exhibit comparable values of $T_c$, indicating that superconductivity is not strongly coupled to electronic localization in normal-state transport.

The coherence length extracted from Fig.~\ref{fig:Hc} provides an explanation for this decoupling. With $\xi(0) \approx \SI{7}{\nano\meter}$ and columnar widths of $\approx\SI{25}{\nano\meter}$ (Fig.~\ref{fig:combined}), the superconducting order parameter is confined well within individual columns rather than averaging across the boundary network. Pairing is therefore set by the local environment of the column interior, while hopping transport is governed by the highest-resistance path through the film, namely the intercolumnar boundaries. $T_0$ and $T_c$ consequently probe spatially distinct regions of the same film. This picture is further consistent with the field-induced broadening evident in Fig.~\ref{fig:Hc}a, since in a granular film the resistive transition marks the onset of phase coherence across the columnar array rather than the loss of pairing within columns.

\section{Conclusion}

In summary, we have demonstrated that IBAS grown ZrN films can span metallic, superconducting, and insulating regimes with basic tuning of growth conditions. Nitrogen partial pressure and deposition temperature act as control parameters, modifying columnar microstructure, sheet resistance, and low-temperature conduction behavior. Superconducting films exhibit strong ZrN(111) texture and type-II behavior, though the measured field dependence departs from the WHH dirty-limit prediction and is better captured by the free-exponent form, yielding $\mu_0 H_{c2}(0) \approx \SI{6.4}{\tesla}$ and $\xi(0) \approx$ \SI{7}{nm}. Increasingly resistive films transition to hopping-dominated transport described by 3D Mott VRH, and the monotonic scaling of the characteristic hopping temperature $T_0$ with sheet resistance confirms disorder-driven localization as the governing mechanism in the normal state. The lack of correlation between $T_0$ and $T_c$ can be explained by the ratio of the coherence length to the columnar width. The two are set by different regions in the film as superconductivity develops within individual columns, whereas hopping is controlled by intercolumnar boundaries. 

These findings position IBAS-grown ZrN as a versatile and controllable quantum material system in which disorder can be engineered without chemical substitution or post-growth processing. The ability to span metallic, localized, and superconducting regimes within a single materials platform makes ZrN attractive for studying disorder-mediated electronic phase evolution and for applications requiring tunable superconducting and resistive properties in CMOS-compatible nitride systems.

\begin{acknowledgments}
The authors gratefully acknowledge Ashley Bielinski for assistance with XRD measurements and analysis, and Genshang Weng for support with dilution refrigerator measurements. The authors also thank Whitney Armstrong and Volodymyr Yefremenko for valuable discussions on thin film growth and electrical transport measurements.

This work was supported by the U. S. Department of Energy (DOE), Office of Science, Offices of Nuclear Physics, Basic Energy Sciences, Microelectronics Initiative, Materials Sciences and Engineering Division under Contract \#DE-AC0206CH11357.
\end{acknowledgments}

\nocite{*}
\bibliography{ZrN_refs.bib}

\end{document}